\documentclass{appolb}

\usepackage{epsfig}

\begin{document}
\title{\bf {Sterile neutrinos and structure formation}%
\thanks{Presented at the XLVII Cracow School of Theoretical Physics, Zakopane, Poland,
June 2007.}%
}
\author{
Jaroslaw Stasielak
\address{Korea Astronomy and Space Science Institute, Daejeon 305-348, Korea, and\\
Institute of Physics, Jagiellonian University, 30-059 Krak\'ow, Poland}
\and
Peter L. Biermann
\address{Max-Planck Institute for Radioastronomy, Bonn, D-53121, Germany,\\
Department of Physics and Astronomy, University of Bonn, D-53121, Germany, and
Department of Physics and Astronomy, University of Alabama, AL 35487, Tuscaloosa, USA}
\and
Alexander Kusenko
\address{Department of Physics and Astronomy, University of California, CA 90095-1547, Los
Angeles, USA}}
\maketitle
\begin{abstract}
Warm dark matter is consistent with the observations of the large-scale
structure, and it can also explain the cored density profiles on smaller
scales. However, it has been argued that warm dark matter could  delay the
star formation.  This does not happen if warm dark matter is made up of keV
sterile neutrinos, which can decay into X-ray photons and active neutrinos.  
The X-ray photons have a catalytic effect on the formation of molecular
hydrogen, the essential cooling ingredient in the primordial gas. 
In all the cases we have examined, the overall effect of sterile dark matter is
to facilitate the cooling of the gas and to reduce the minimal mass of the halo
prone to collapse.  We find that the X-rays from the decay of keV sterile
neutrinos facilitate the collapse of the gas clouds and the subsequent star
formation at high redshift.
\end{abstract}
\section{Introduction}

Both cold and warm dark matter models agree with the observed structure 
on the large scales.  However, there are several inconsistencies between the
predictions of the cold dark matter (CDM) model and the
observations~\cite{cdm_problems}.   The low cutoff in dark matter contents of
dwarf spheroids, the smoothness of our dark matter halo, and the old globular
clusters (observed in Fornax) resisting the infall into the center by dynamical
friction~\cite{cdm_problems}, all can be explained by warm dark matter
(WDM) because it suppresses the structure on scales that are smaller
than the free-streaming length. 

While the suppression of the small-scale structure is desirable, it has been
argued that ``generic'' WDM (for example, gravitino) can slow down structure
formation and delay reionization of the 
universe, which can lead, in turn, to an inconsistency with the reionization
redshift obtained by the WMAP \cite{Yoshida}.  
This problem can be alleviated in the case of the WDM  in
the form of sterile neutrinos with mass of several keV and a small mixing angle
with the ordinary neutrino \cite{bier,stas,stas07} because such sterile
neutrinos can decay and produce photons that catalyze the formation molecular
hydrogen and speed up the star formation.  

In the absence of metals, gas cooling is mainly due to
the collisional excitation of H$_{2}$, its subsequent spontaneous de-excitation,
and photon emission. In the primordial gas clouds, hydrogen molecules can be
formed only in reactions involving $e^{-}$ or H$^+$ as a catalyst. Thus, an
X-ray radiation can increase the production of the H$_{2}$ by enhancing the
ionization fraction, which subsequently leads to speed up of the gas cooling and
star formation.
Although sterile neutrinos are stable on cosmological time scales, they nevertheless decay. The decay channel important for us is that of decay into one active neutrino and one photon, i.e., $\nu_s \rightarrow \nu_a \gamma$, where the photon energy is half of the sterile neutrino mass, $E_{0} \approx m_{s}c^2/2$. 
These decays produce an X-ray background radiation that increases the production
of molecular hydrogen and can induce a rapid and prompt star formation at high
redshift.

Sterile dark matter has a firm motivation from particle
physics~\cite{dw,Kusenko:2006rh}.  The discovery
of the neutrino masses implies the existence of right-handed gauge-singlet
fields, all or some of which can be lighter than the electroweak scale.  These
sterile neutrinos can be produced in the early universe by different
mechanisms, for example, from neutrino oscillations~\cite{dw} or from the Higgs
decays~\cite{Kusenko:2006rh}, or from the couplings to a low-scale 
inflaton~\cite{Shaposhnikov:2006xi}.   The same particles, produced in a
supernova, could account for the supernova asymmetries and the pulsar
kicks~\cite{kus}, and can play a role in the formation of super-massive black
holes in the early universe~\cite{puzzle}.

We will examine the thermal evolution of the gas clouds, taking into
account both effects of the sterile neutrino decays,
namely, the ionization and heating of the gas. 
We follow the evolution of the baryonic top-hat overdensity, the gas
temperature and the H$_{2}$ and $e^{-}$ fraction. 
In order to perform the calculation we have incorporated to our previous code \cite{stas}, the effects of sterile neutrino decays within collapsing halos and absorption of the X-ray background from sterile neutrinos by He atoms in the intergalactic medium. Our goal is to juxtapose
the evolution of the gas temperature in the primordial clouds in the 
CDM model and the WDM model with keV sterile neutrinos and estimate of the
minimal mass able to collapse at a given redshift. 
%
%
%

\section{Description of the code}
\label{ana}

The top-hat overdensity evolution in a single-zone approximation \cite{tegmark,stas} is described by the following equation \cite{Padmanabhan}
\begin{equation}
	\delta=\frac{9}{2}\frac{\left(\alpha-\sin
\alpha\right)^{2}}{\left(1-\cos \alpha\right)^{3}}-1 \textrm{,} \label{th1}
\end{equation}
where the parameter $\alpha$ is related to the redshift $z$ and the redshift of
virialization $z_{vir}$ through
\begin{equation}
	\frac{1+z_{vir}}{1+z}=\left(\frac{\alpha- \sin \alpha}{2
\pi}\right)^{2/3}. \label{th2}
\end{equation}
According to these equations, the virial value of overdensity ($\delta \approx 18 \pi^2$) is reached at the redshift $z_{3\pi/2}=1.06555\left(1+z_{vir}\right)-1$.

Further evolution of $\delta$ depends on the type of the matter in the overdense 
region. If it is the dark matter, then after virialization its density remains 
constant forever. The situation is different in the case of baryons. If cooling is efficient enough, then the density gradually
increases. Otherwise, the density remains constant, and there is no star
formation in the halo. 
Following \cite{tegmark}, we assume that the density of the halo stays
constant after redshift of $z_{3\pi/2}$, i.e., it is equal to the value $\rho = 18\pi^{2}\Omega_{0}\rho_{0}\left(1+z_{vir}\right)^{3}$,
where $\rho_{0}$ is the present critical density of the universe. 
This assumption is sufficient for our purposes.

The evolution of the gas temperature is governed by the equation \cite{stas}
\begin{eqnarray}
\frac{dT}{dz}&=& \left(\gamma-1\right)\frac{T}{n_{p}}\frac{dn_{p}}{dz}+\gamma\frac
{T}{\mu}\frac{d\mu}{dz}+\frac{T}{\left(\gamma-1\right)}\frac{d\gamma}{dz} \nonumber \\
&+& 	\frac{\left(\gamma-1\right)\Lambda}{n_{p}k
H_{0}\left(1+z\right)\sqrt{\Omega_{\Lambda}+\Omega_{0}\left(1+z\right)^{3}}}
\textrm{,} \label{tz}
\end{eqnarray}
where $T$, $\gamma$, $n_{p}$, $\mu$, $k$ and $\Lambda$ are the temperature of the gas, the adiabatic index, the number density of non-dark matter particles, the molecular weight, the Boltzmann constant and the cooling/heating function, respectively. 

If we simply integrate equation (\ref{tz}) to the redshift $z_{vir}$ then, for primordial clouds with masses greater than a critical value, the gas temperature will be much lower than the virial temperature $T_{vir}$~\cite{Barkana} 
\begin{equation}	
T_{vir}\approx 12.3\times10^{3}\textrm{ K
}\left(\frac{\mu_{vir}}{0.6}\right)\left(\frac{M}{10^{8}h^{-1}M_{\odot}}
\right)^{2/3}  \left( \frac{1+z_{vir}}{10} \right)
  \textrm{,} \label{virial}
\end{equation}
where $\mu_{vir}$ and $M$ are the mean molecular weight during virialization and the halo mass (combined mass of dark and baryonic constituents).
Therefore, we must take into account shocks and increase the gas temperature
to the virial value. We assume that the evolution of the gas temperature
is linear between the redshifts $z_{3\pi/2}$ and $z_{vir}$.

To calculate the cooling/heating function $\Lambda$ in equation (\ref{tz}) we follow the number density evolution of different baryonic components and take into account all relevant chemical and thermal processes, e.g, heating and cooling due to the sterile neutrino decays. 
The detailed list of all included processes and their coefficients can be found in \cite{stas}.

Let us define the number fraction of component $i$ as $	x_{i}=n_{i} / n$, where $n_{i}$ and $n=n_{H}+n_{H^{+}}+2n_{H_{2}}$ are the number density of the $i$th component and hydrogen species, respectively.
The time evolution of number fraction can be described by the kinetic equation:
\begin{equation}
{dx_i \over dt} = n \sum_{l} \sum_{m} k_{lmi} x_l x_m +
\sum_{j} k_{ji} x_j \textrm{,}
\label{chemia}
\end{equation}
where the first component on the right-hand side describes the chemical reactions and the other one accounts for photoionization/photodissociation processes. Coefficients
$k_{lmi}$ and $k_{ji}$ are reaction rates and photoionization/photodissociation rates multiplied by the numbers equal to 0, $\pm 1$ or $\pm 2$
depending on the reaction. In our calculations, we have considered the following five species: H, H$^{+}$, ${\rm H}_{2}$, $e^{-}$, and $\rm He$ with the mass fraction $Y=0.244$ \cite{Izotov}. We used simplified molecular hydrogen chemistry similar to the approach presented in \cite{tegmark}.

In order to take into account the effects of sterile neutrino decays inside the collapsing halo, we have solved the radiative transfer equation for the spherically symmetric clouds with uniform density (see Section \ref{rad} and also \cite{stasielak05,mihalas}). In addition, we have included absorption of the X-rays from the sterile neutrino decays by both $H$ and $He$. 

The photons from the decay of the sterile neutrinos are mainly absorbed by
neutral helium and hydrogen atoms leading to their ionization. The ionization rate due to
these photons is enhanced almost 100 times due to additional ionization by the
secondary electrons, which deposit almost 1/3 of their energy into ionization. The
energy of the absorbed photons partially goes into ionization and partially
into heating and excitations. We have adopted the approximation
\cite{steen,stas}, in which the ionization rate (in units s$^{-1}$) and heating (in units erg s$^{-1}$ cm$^{-3}$) due to the photons from
the decay of the sterile neutrinos are respectively equal to
\begin{eqnarray}
k\left(z\right) &=& \left[ \int_{\nu^{H}_{th}}^{\infty}4\pi  
\sigma_{H}\left(\nu \right)  
\frac{I_{\nu}\left(z\right)}{h\nu} \left(\frac{h \nu -h \nu_{th}^{H}}{h
\nu_{th}^{H}} \right) d\nu \right. \nonumber \\
&+& \left. \frac{Y}{4X} \int_{\nu^{He}_{th}}^{\infty}4\pi  
\sigma_{He}\left(\nu \right)  
\frac{I_{\nu}\left(z\right)}{h\nu} \left(\frac{h \nu -h \nu_{th}^{He}}{h
\nu_{th}^{H}} \right) d\nu +\frac{\Lambda_{int}\left(z\right)}{h \nu_{th}^H n_{H}} \right] \nonumber \\
&\times& C_{i}\left(1-x_{e}^{a_{i}}\right)^b_{i} +
\int_{\nu_{th}^{H}}^{\infty}4\pi\sigma_{H}\left(\nu\right)\frac{I_{\nu}
\left(z\right)}{h\nu} d\nu \label{ion} \textrm{,}
\end{eqnarray}
\begin{eqnarray}
\Gamma_{s}\left(z\right) &=&  \left[ \int_{\nu^{H}_{th}}^{\infty}4\pi  
\sigma_{H}\left(\nu \right)  
\frac{I_{\nu}\left(z\right)}{h\nu} \left(h \nu -h \nu_{th}^{H}  \right) d\nu
\right. \nonumber \\
&+& \left. \frac{Y}{4X} \int_{\nu^{He}_{th}}^{\infty}4\pi  
\sigma_{He}\left(\nu \right)  
\frac{I_{\nu}\left(z\right)}{h\nu} \left(h \nu -h \nu_{th}^{He}  \right) d\nu
+\frac{\Lambda_{int}\left(z\right)}{n_{H}}\right] \nonumber \\
&\times& C_{h}\left[ 1-\left(1-x_{e}^{a_{h}}\right)^b_{h} \right] n_{H}
\label{heat} \textrm{,}
\end{eqnarray}
where $h$, $\nu$, $\sigma_{H}\left(\nu \right)$, $h \nu_{th}^{H}=13.6$ eV, $\sigma_{He} \left( \nu \right)$, $h \nu_{th}^{He}=24.6$ eV and $X$ 
are the Planck constant, photon frequency, the cross section and energy threshold for $H$ and $He$ ionization, and hydrogen mass ratio, respectively. 
The coefficients
$C_{i}=0.3908$, $a_{i}=0.4092$,
$b_{i}=1.7592$,~$C_{h}=0.9971$, $a_{h}=0.2663$ and $b_{h}=1.3163$ are taken from \cite{steen}.
The function $\Lambda_{int}\left(z\right)$ (in units erg s$^{-1}$ cm$^{-3}$) is
 the energy absorption rate of the photons from the sterile neutrino decays inside the collapsing cloud and is given by equation (\ref{lambdainsidefin}). Finally, $I_{\nu}\left(z\right)$ (in units of erg cm$^{-2}$ s$^{-1}$ sr$^{-1}$ Hz$^{-1}$) is the specific intensity of the X-ray background from the sterile neutrino decays, which takes into account absorption by H and He in the intergalactic medium and which can be calculated in a similar way to the specific intensity given in \cite{stas}. 
 
Since the cross section for absorption of the X-ray photons by He is much larger
than the same cross section for H,  a large amount of energy can be accumulated
in free electrons due to He ionization. Therefore, changes in the He ionization
can strongly affect the absorbed energy, and we  must take into account
their effects to the heating/cooling function and the ionization by the
secondary electrons, as in equations (\ref{ion}) and (\ref{heat}).

To estimate the minimal mass of the primordial halo able to
collapse at a given redshift we use the following criterion of successful
collapse~\cite{tegmark}:
\begin{equation}
	T \left( \eta z_{vir} \right) \leq \eta T \left(z_{vir} \right) \label{collaps} \textrm{,}
\end{equation}
where we take $\eta=0.75$. It means that the cloud is considered to collapse if its temperature drops substantially
within a Hubble time, which 
roughly corresponds to
the redshift dropping by a factor $2^{2/3}$.

\section{Radiative transfer inside the collapsing cloud}
\label{rad}
In order to derive
the energy absorption rate of the photons from the sterile neutrino decays inside the collapsing gas clouds $\Lambda_{int}\left(z\right)$,
let us assume that these halos have spherically symmetric shape and their densities are uniform. In that case,
it will be convenient to consider the radiative transport equation in a system
of coordinates $\left(r, p\right)$ defined by the transformation formula:
$\left(r, \mu \right) \rightarrow \left(r, p = r\sqrt{1-\mu^{2} }\right)$ for
$-1\leq \mu \leq 1$ \cite{hu71,stasielak05}, 
where $r$ is a radial coordinate, $\mu = \cos \theta$, and $\theta$ is the angle
between the outward normal and the photon direction. 

For a given radius $r$, the "impact" parameter $p$ can vary between 0 and $r$. Because the parameter $p$ cannot distinguish between $\mu >0$ and $\mu < 0$, the radiation intensity $I_{\nu}$ has to be separated into outward $I^{+}_{\nu}$ and inward $I^{-}_{\nu}$ directed intensity, respectively.

Now, we can cast the time-independent, non-relativistic equation for radiation transport in spherical geometry into the form
\begin{eqnarray}
	\frac{\partial j_{\nu} \left(\tau_{\nu},p\right)}{\partial \tau_{\nu}}
&=& h_{\nu}\left(\tau_{\nu},p\right) \label{jjj-i}
\label{eqq1} \\
	\frac{\partial h_{\nu} \left(\tau_{\nu},p\right)}{\partial \tau_{\nu}}
&=& j_{\nu}\left(\tau_{\nu},p\right)-S_{\nu} \
\textrm{,} \label{eqq2}
\end{eqnarray}
where
\begin{eqnarray}
	j_{\nu} \left(r,p\right) &=& \frac{1}{2} \left(   I_{\nu}^{+}
\left(r,p\right) + I_{\nu}^{-} \left(r,p\right) \right) \label{fe1} \textrm{,}
\\
	h_{\nu} \left(r,p\right) &=& \frac{1}{2} \left(   I_{\nu}^{+}
\left(r,p\right) - I_{\nu}^{-} \left(r,p\right) \right) \textrm{,} \label{fe2} 
\end{eqnarray}
and $d \tau_{\nu} = - \chi_{\nu}  d\left( r \mu \right)$, is the optical depth at the radius $r$. The term 	$S_{\nu} = \eta_{\nu}/\chi_{\nu}$ is a source function, where $\eta_{\nu}$ (in units of erg s$^{-1}$ cm$^{-3}$ sr$^{-1}$ Hz$^{-1}$) and $\chi_{\nu}$ (in units of cm$^{-1}$) are the total emissivity and opacity at frequency $\nu$, respectively.

We consider sterile neutrino decays inside the spherically symmetric cloud, thus, the photon flux from this process will be peaked around the frequency corresponding to the energy of $E_{0}=m_{s}c^2/2$. Since the density of the cloud is much higher than its surroundings, there will be only few photons at this energy impinging upon the outer boundary, which is set by the radius $R$ of the cloud. The external radiation field also consists of the X-ray photons emitted due to the 
sterile neutrino decays at large distances from the cloud, however, as they have been emitted at earlier times, they will be redshifted to lower energies. Thus, they will not give the contribution to the photon flux at the energy of $E_{0}$.

Assuming, that there is no external radiation field at the energy of $E_{0}$, we can write the boundary conditions as follows
\cite{stasielak05}
\begin{eqnarray}
	h_{\nu} \left(p,p\right) &=& 0 \hskip 1.5cm 0 \leq p \leq R  \label{con1} \textrm{,} \\	
	j_{\nu} \left(R, p \right) &=& h_{\nu}\left(R,p\right) \hskip 1cm 0<p<R  \textrm{.}\label{w2}
\end{eqnarray}

Since we are interested only in the photon flux at the energy of $E_{0}$, from now on, we drop the $\nu$ dependence for clarity. It means that we have to multiply all of the quantities by the Dirac delta $\delta \left(E_{0}/h\right)$ or understand them as they  have been already integrated over frequency. We will use the latter interpretation. According to the emissivity and opacity definition, we have
\begin{eqnarray}
	\eta &=& \frac{d  n_{\gamma}}{d t} \frac{E_{0}}{4\pi } = \frac{\Omega_{dm} c^2 \varrho }{8\pi \tau_{s}}  \textrm{,} \label{kap22} \\
	\chi &=& n_{H}\sigma_{H} \left(E_{0}\right)+ n_{He}\sigma_{He}\left(E_{0}\right)  \textrm{,} \label{kap1}
\end{eqnarray}
where $\varrho$ is the total dark and baryonic matter mean density of the collapsing cloud, $n_{\gamma}$ is the number density of emitted photons due to the sterile neutrino decays. 
We assume that all of the dark matter in the collapsing halo consists of the sterile neutrinos and that the number density of sterile neutrinos do not change with time. The latter assumption can be justified by the fact that
the inverse width of the sterile neutrino radiative decay, $\tau_{s}$, is much longer than the age of the universe. In addition, we neglect the stimulated emission.

Equations (\ref{eqq1}) and (\ref{eqq2})
can be rewritten as
\begin{eqnarray}
	\frac{\partial^2 j \left(\tau,p\right)}{\partial \tau^2} &=& j \left(\tau,p\right)-S \textrm{,} \\
	\frac{\partial^2 h \left(\tau,p\right)}{\partial \tau^2} &=& h \left(\tau,p\right)  \textrm{,}
\end{eqnarray}
which with the boundary conditions (\ref{con1}) and (\ref{w2}) have the following solution
\begin{eqnarray}
	h\left(\tau,p\right) &=& -S e^{T\left(p\right)}  \sinh \tau \textrm{,} \label{hhhh} \\
	j\left(\tau,p\right) &=& -S e^{T\left(p\right)} \cosh \tau +S \textrm{,}
\end{eqnarray}
where $T\left(p\right)=-\chi  R \mu$ is the optical depth at the cloud boundary derived for the given impact parameter $p$.
The luminosity of the collapsing cloud is given by
\begin{equation}
	L = 16 \pi^{2} R^{2} \int_{0}^{1} h \left( R,\mu \right) \mu d\mu  \label{luminosity} \textrm{,}
\end{equation}
whereas if we neglect absorption it would be equal to
\begin{equation}
	L_{s} = \frac{16}{3} \pi^{2} R^{3} \eta   \label{luminositys} \textrm{.}
\end{equation}
The fraction of the energy absorbed by the cloud is given by $1-L/L_{s}$, thus the energy absorption rate (in units of erg cm$^{-3}$ s$^{-1}$) is equal to
\begin{equation}
	\Lambda_{int}=\left( 1-\frac{L}{L_{s}}\right) L_{s} \frac{3}{4\pi R^3} \label{lambdainside} \textrm{.}
\end{equation}

Using equations (\ref{kap22}), (\ref{kap1}), (\ref{hhhh}), (\ref{luminosity}) - (\ref{lambdainside}), the definition of the source function $S$, and doing some algebra we get
\begin{equation}
	\Lambda_{int}\left(z\right)= \frac{\Omega_{dm} m_{H} c^2}{2\Omega_{b}\tau_{s} X} f \left( \alpha \right)  n\left(z\right) \label{lambdainsidefin} \textrm{,}
\end{equation}
where
\begin{eqnarray}
	f\left( \alpha \right) &=& 1- \frac{3}{4}\frac{1}{\alpha^3}\left[\alpha^2-\frac{1}{2}+\left(\alpha+\frac{1}{2}\right) e^{-2\alpha}\right] \textrm{,} \label{lll} \\
	\alpha \left( z \right) &=& \left(  \frac{3 \pi \Omega_{b} X M}{4 m_{H} n\left(z\right)}   \right)^{1/3}
	\left[ n_{H}\left(z\right) \sigma_{H}\left(E_{0}\right)+ n_{He}\left(z\right) \sigma_{He}\left(E_{0}\right) \right]
\textrm{.}
\end{eqnarray}
The term $m_{H}$ and $M$ denotes hydrogen mass and the total mass of the cloud, respectively.

\section{Results}
\label{res}

\begin{figure}[t]
	\centering
		\includegraphics[width=1.00\textwidth]{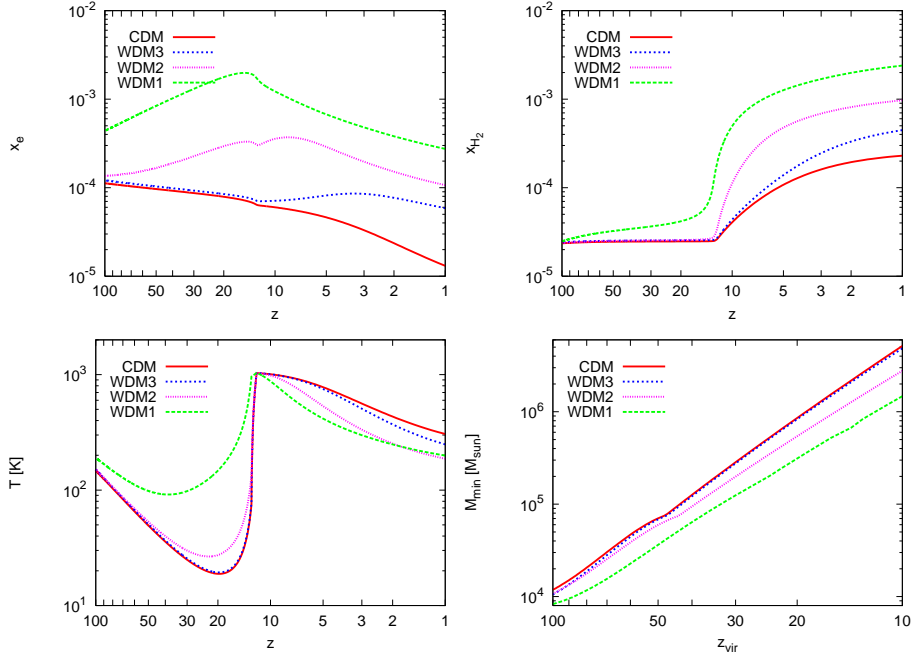}
	\caption{\textit{Top left}, \textit{top right} and \textit{bottom left}: Evolution of ionization fraction, $H_{2}$ fraction and temperature with redshift for different models. In each case, the mass of the primordial cloud is equal to $M=10^6 M_{\odot}$ and virialization redshift to $z_{vir}=12$. 
\textit{Bottom right}: Dependence of the minimal mass of primordial halo able to collapse on its virialization redshift. Models we have used in calculation are following: $m_{s}=25$ keV and $\sin^2 \theta=3 \times 10^{-12}$ (WDM1), $m_{s}=15$ keV and $\sin^2 \theta=3 \times 10^{-12}$ (WDM2), $m_{s}=3.3$ keV and $\sin^2 \theta=3 \times 10^{-9}$ (WDM3), and CDM.}
	\label{fig:f1}
\end{figure}

We have performed a detailed analysis of the cooling and collapse of the primordial
gas clouds in the model with warm dark matter, taking into account both the increase
in the fraction of molecular hydrogen and the heating due to
the sterile neutrino decays. To illustrate these effects, we have performed the analysis for some benchmark cases which arise in realistic scenarios \cite{stas}. 
The effect on the largest gas clouds
is negligible, whereas smaller clouds will be affected: for
the largest clouds, the additional molecular hydrogen makes no difference, but for the 
smaller ones, the increase in the X-ray background makes the collapse possible in cases
where it could not occur in the absence of sterile neutrino decays. 

Our results presented in Fig.~\ref{fig:f1} show that  the
overall effect of sterile neutrino decays is to enhance the $H_{2}$ fraction and
to speed up the cooling of the gas in the primordial halos. The minimal mass of
the cloud able to collapse is reduced in all WDM models we have examined. 
We note that a more detailed treatment of the sterile neutrino free-streaming
may affect our results: we did not take into account
the filamentary star formation~\cite{Gao:2007yk}; some other effects
may also be important~\cite{ripamonti}. 

In summary, the X-ray photons from sterile neutrino decays could play an
important role in the formation of the first stars because they increase 
the fraction of molecular hydrogen.

\end{document}